\documentclass[sigconf]{acmart}

\AtBeginDocument{%
  \providecommand\BibTeX{{%
    \normalfont B\kern-0.5em{\scshape i\kern-0.25em b}\kern-0.8em\TeX}}}

\usepackage{dblfloatfix}
\usepackage{xcolor}
\usepackage{enumitem}
\setlist{nosep}
\usepackage{graphicx}
\usepackage{algpseudocode} 
\usepackage[ruled,linesnumbered]{algorithm2e}
\usepackage{subcaption}
\usepackage{balance}
\usepackage{tabularx}

\usepackage{titlesec}
\titlespacing*{\section}
{0pt}{3ex}{1ex}
\titleformat{\section}
{\normalfont\Large\bfseries\MakeUppercase}{\thesection}{1em}{}

\titlespacing*{\subsection}
{0pt}{2ex}{1ex}

\definecolor{darkblue}{rgb}{0,0,0.5}

\definecolor{mygray}{gray}{0.9}

\newcommand{\laterurl}[1]{\href{#1}{this link}}

\renewcommand\footnotetextcopyrightpermission[1]{}

\begin{document}

\title{Efficient Real-Time Selective Genome Sequencing on Resource-Constrained Devices}

\author{Po Jui Shih$^1$, Hassaan Saadat$^1$, Sri Parameswaran$^1$, Hasindu Gamaarachchi$^{2,1}$}
\affiliation{
    \institution{$^1$School of Computer Science and Engineering, UNSW Sydney, Australia}
}
\affiliation{
    \institution{$^2$Kinghorn Centre for Clinical Genomics, Garvan Institute of Medical Research, Sydney, Australia}
}



\begin{abstract}
Third-generation nanopore sequencers offer a feature called selective sequencing or `Read Until' that allows genomic reads to be analyzed in real-time and abandoned halfway, if not belonging to a genomic region of `interest'. This selective sequencing opens the door to important applications such as rapid and low-cost genetic tests. The latency in analyzing should be as low as possible for  selective sequencing to be effective so that  unnecessary reads can be rejected as early as possible. However, existing methods that employ subsequence Dynamic Time Warping (sDTW) algorithm for this problem are too computationally intensive that a massive workstation with dozens of CPU cores still struggles to keep up with the data rate of a mobile phone-sized MinION sequencer. In this paper, we present Hardware Accelerated Read Until (HARU), a resource-efficient hardware-software co-design-based method that exploits a low-cost and portable heterogeneous MPSoC platform with on-chip FPGA to accelerate the sDTW-based Read Until algorithm. Experimental results show that HARU on a Xilinx FPGA embedded with a 4-core ARM processor is around 2.5$\times$ faster than a highly optimized multi-threaded software version (around 85$\times$ faster than the existing unoptimized multi-threaded software) running on a sophisticated server with 36-core Intel Xeon processor for a SARS-CoV-2 dataset. The energy consumption of HARU is two orders of magnitudes lower than the same application executing on the 36-core server. Source code for HARU sDTW module is available as open-source at \url{https://github.com/beebdev/HARU} and an example application that utilises HARU is at \url{https://github.com/beebdev/sigfish-haru}.

\end{abstract}

\maketitle
\pagestyle{plain}

\section{Introduction} \label{s:introduction}

\setlength{\textfloatsep}{2ex}
\begin{figure}[t] 
    \centering
    \includegraphics[width=0.9\linewidth,trim={8.7cm 7.5cm 8.5cm 2cm},clip]{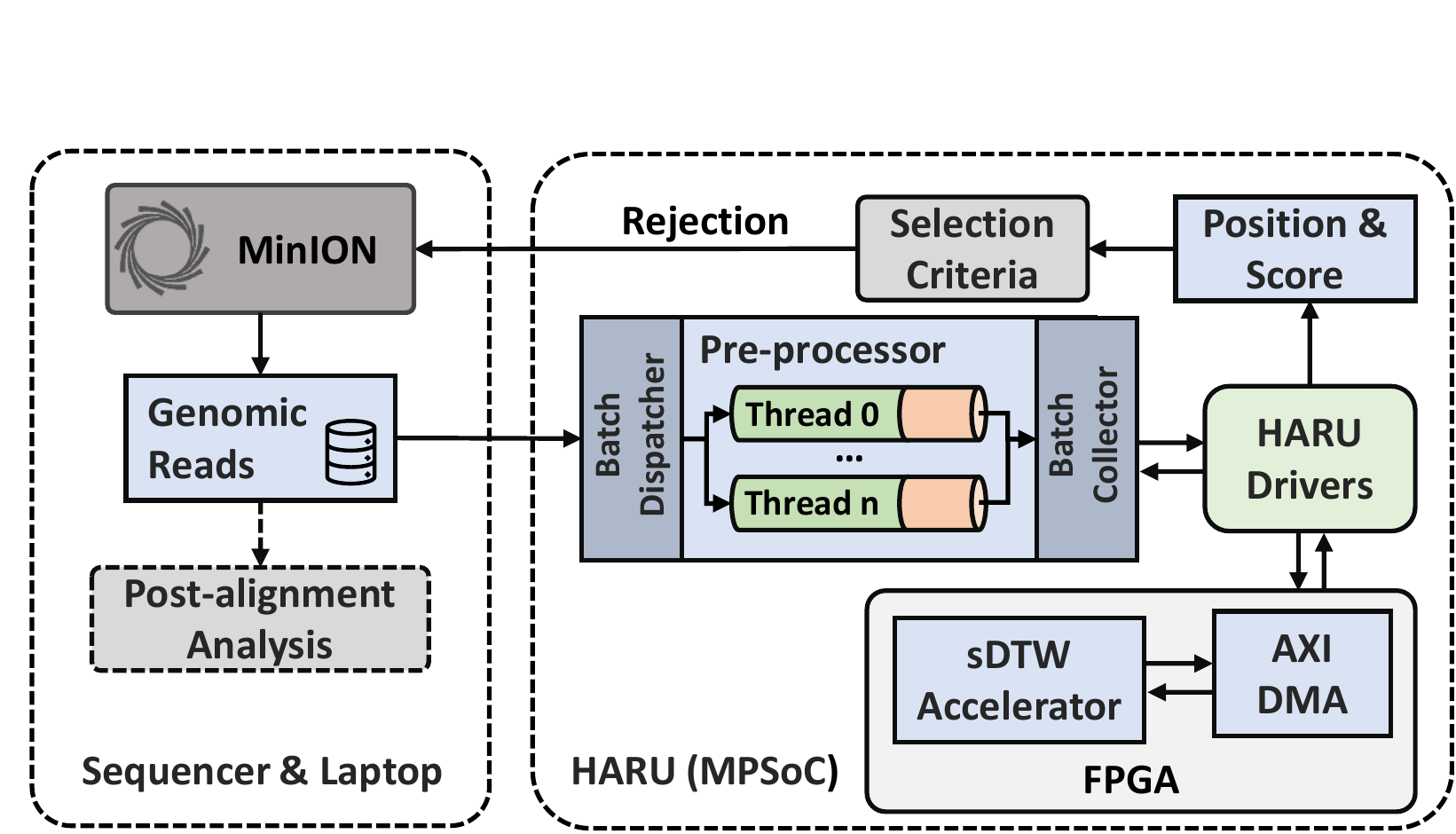}
    \setlength{\abovecaptionskip}{1.7ex}
    \caption{HARU overview.} \label{f:haru_overview}
\end{figure}

Nanopore sequencers belonging to the latest (third) generation sequencing technology have revolutionized the field of genomics. The portable palm-sized nanopore sequencer called the MinION produced by Oxford Nanopore Technologies (ONT) is capable of performing direct selective sequencing which rejects the genomic reads that are not of interest. This technique, also known as Read Until, can vastly reduce the sequencing time and cost for applications such as genetic disease identification \cite{Stevanovski2022,miller2021targeted}, cancer detection \cite{djirackor2021intraoperative,yamaguchi2021application}, the surveillance of viruses (e.g. SARS-CoV-2) and other pathogens \cite{wang2020nanopore,marquet2022evaluation}, and sequencing low abundance species metagenomics samples \cite{martin2022nanopore}. However, the real-time analysis of genomic reads which involves aligning the read to the reference to obtain the position information is a complex and time-consuming process. Existing alignment methods used in selective sequencing either directly map the sequenced raw signals using algorithms such as the subsequence dynamic time warping (sDTW) \cite{loose2016real} or convert the raw signals into base sequenced before using base-domain alignment methods \cite{edwards2019real, payne2020nanopore}.

Previous sDTW-based Read Until running on a 22-core High-Performance Computing (HPC) system \cite{loose2016real} could not keep up with the current 450 bases/s sequencing rate \cite{wang2021nanopore} of the portable palm-sized MinION sequencer and we found that the sDTW computation takes more than 98\% of its run time. Current base-aligning Read Until implementations \cite{payne2020nanopore} use GPU-accelerated basecallers to convert signal reads to bases and map with the reference base sequence using optimized sequence mapping techniques (e.g. \textit{Minimap2} \cite{li2018minimap2}) that are mature in the bio-informatics field. Although enough to keep up with the sequencing rate, the execution requires high-end GPU hardware (NVIDIA RTX 1080 for simple reference targets \cite{payne2020nanopore} and NVIDIA RTX 3090 for more complex targets \cite{Stevanovski2022}) which makes selective sequencing expensive and non-portable.

To address the lack of portability and costly execution nature of existing implementations, we return to raw signal mapping Read Until and tackle the computational bottleneck by accelerating the sDTW algorithm with FPGAs. Hardware acceleration work in the past for the sDTW algorithm is rare and the $O(MN)$\footnote{$M$ is the size of the query and $N$ is the size of the reference sequence} space complexity of the algorithm does not naively fit on reconfigurable hardware efficiently. For the use case of selective sequencing, the $M\times N$ memory is unnecessary and required results can be obtained with $O(M)$ space complexity. We designed a memory-efficient sDTW accelerator for Read Until and exploited fine-grain parallelism to reduce the computational time complexity to $O(M+N)$. 

In this paper, we propose HARU (see Figure \ref{f:haru_overview}), a software-hardware co-design for raw signal-alignment Read Until that utilizes the memory-efficient sDTW accelerator for high throughput signal mapping. HARU targets low-cost heterogeneous MPSoC devices that have on-chip reconfigurable hardware and performs efficient multi-threaded batch-processing for the signal preparation in conjunction with the sDTW accelerator. We demonstrate that our design is capable of keeping up with the data rate of the ONT's MinION sequencer and gains around 85$\times$ speedup against the original software implementation mapping the SARS-CoV-2 sequenced data on a 36-core HPC system. Furthermore, we show that our work runs around 2.5$\times$ faster than an optimized multi-threaded software implementation on the same 36-core server and around 6.5$\times$ faster than the same software running on a 10-core Intel Core i9-10850K desktop. Multiple instances of the system running on low-cost MPSoC can be parallelized to scale up processing and run on sequencers of a larger scale (e.g. ONT's PromethION). 

The main contributions of our work include:

\begin{itemize}[leftmargin=*]
\item a fully working hardware accelerated signal-matching Read Until designed to perform high throughput sequence mapping on resource-constrained embedded platforms.
\item a resource-efficient sDTW accelerator for selective sequencing that reduces space complexity from $O(MN)$ to $O(M)$ while exploiting fine-grain parallelism to reduce the computational time from $O(MN)$ in software to $O(M+N)$ in hardware.
\item the full proposed design including the software processing layer, devices drivers, and the hardware sDTW accelerator developed during the span of this research is released as open-source at \url{https://github.com/beebdev/HARU} and an example application that utilises HARU is at \url{https://github.com/beebdev/sigfish-haru}.
\end{itemize}

\section{Background} \label{s:background}
\subsection{Nanopore Sequencing} \label{ss:bg-nanopore_seq}

The process of obtaining the genome\footnote{A Genome is a sequence of base-pairs (bp) formed by a quaternary system containing nucleobases: adenine (\textbf{A}), cytosine (\textbf{C}), guanine (\textbf{G}), and thymine (\textbf{T}).} of an organism in computer-readable form is called genome sequencing \cite{heather2016sequence}.
Nanopore sequencers from ONT are third-generation genomic sequencers that are capable of producing long reads\footnote{At the sample preparation stage, the DNA or RNA molecule of the target sample breaks into small fragments of molecules and these fragments are called \textit{reads}.} (currently ranging between 1 kilo-bases to 2 mega-bases) \cite{jain2018nanopore,deamer2016three} and are commercially available at an affordable price compared to sequencers of other techniques and generations \cite{petersen2019third}. These ONT nanopore sequencers provide genomic reads through \textit{flow cells} which contain a proprietary sensor array over nanopore channels embedded on a synthetic membrane \cite{logsdon2020long}. During the sequencing process, the nanopore channels capture the electric current change caused by the genome molecules' ionic current when it passes through \cite{logsdon2020long}. This current signal trace is streamed to the sequencer software in real-time and can later be basecalled\footnote{Basecalling is the process of converting raw signals into a sequence of nucleobases (ACGT).} into the corresponding nucleobase representation for later analysis \cite{wick2019performance}.

\subsection{Selective sequencing} \label{ss:bg-sel_seq}

\setlength{\textfloatsep}{4ex}
\begin{figure}[t] 
    \centering
    \begin{subfigure}[!ht]{\linewidth}
        \centering
        \includegraphics[width=\linewidth,trim={7cm 6cm 7.5cm 5cm},clip]{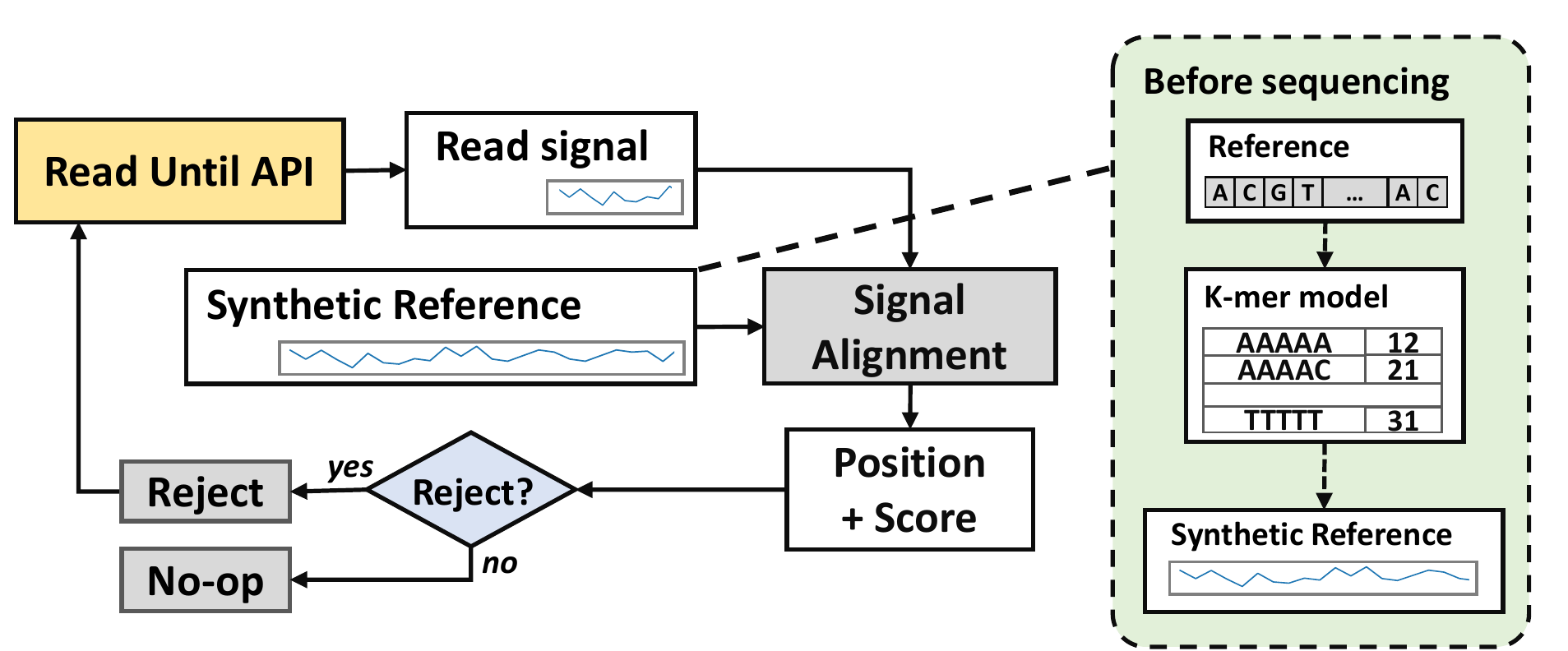}
        \caption{Signal-alignment Read Until} \label{f:signal_ru}
    \end{subfigure}\\
    \begin{subfigure}[!ht]{\linewidth}
        \centering
        \setlength{\belowcaptionskip}{-1ex}
        \includegraphics[width=\linewidth,trim={7.5cm 7cm 7.5cm 5.5cm},clip]{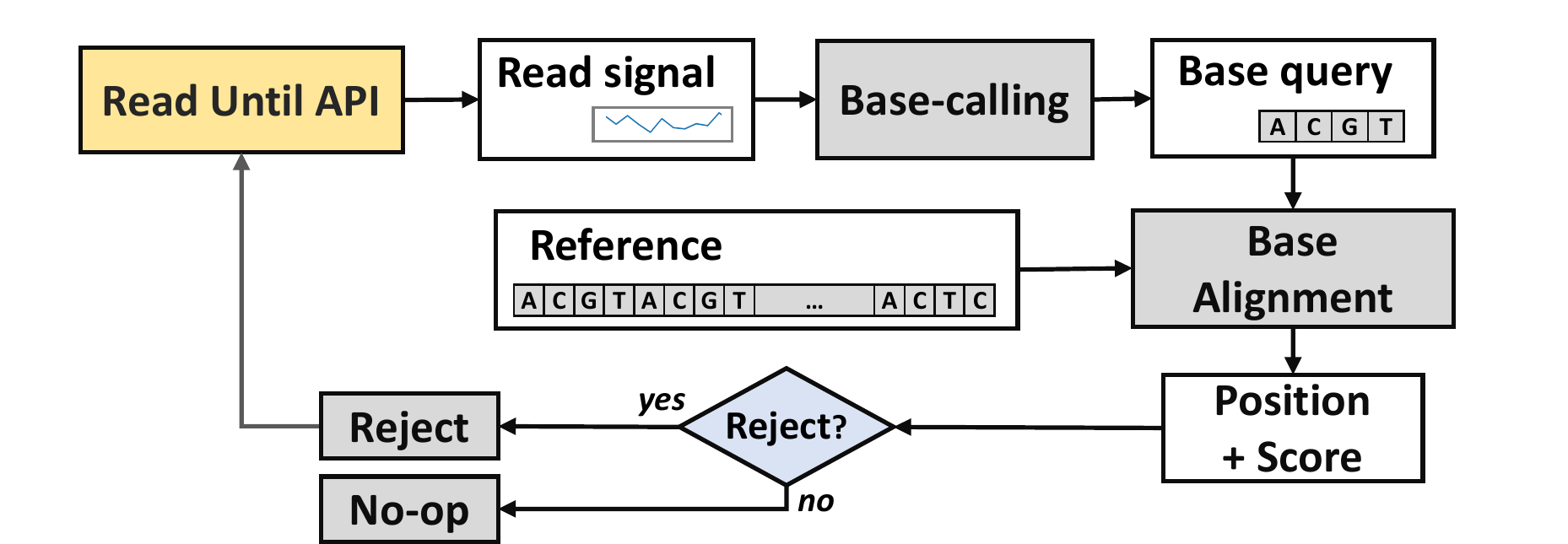}
        \caption{Base-alignment Read Until} \label{f:base_ru}
    \end{subfigure}
    \setlength{\belowcaptionskip}{-3ex}
    \caption{Overview of Read Until workflows}
    \label{f:ru_workflow}
\end{figure}

A feature of ONT nanopore sequencers is the direct selective sequencing capability. These sequencers provide real-time data output stream and allow rejection of reads at individual nanopore channels \cite{loose2016real,payne2020nanopore}. This means the sequenced data can be analyzed during the sequencing and rejected before completion if decided it is not of interest. This selective sequencing process in the nanopore sequencing workflow is known as \textit{Read Until} and ONT provides the Read Until API interface for software applications to access and reject the sequenced reads in real-time. A rejection made through the API call will eventually be passed back to the sequencer and the voltage at the indicated channel will be reversed to eject the genomic molecule out of the nanopore \cite{loose2016real}.

For the Read Until execution to be effective, the round-trip task latency for read acquisition, analysis, and rejection signal forwarding should be completed before the majority of the subject read is sequenced by the nanopore sequencer \cite{loose2016real}. Rejections that are made after the majority of the strand is sequenced bring no benefit as no sequencing time is saved. Existing Read Until methods perform analysis by aligning the genomic reads to the target reference and making the rejection decision based on the position and distance score. This alignment can be done using either signal or base alignment  \cite{loose2016real,zhang2021real,kovaka2021targeted,payne2020nanopore,edwards2019real}.

\textbf{Signal-alignment Read Until.}
Signal-alignment Read Until aligns raw signal reads with the reference to obtain the alignment position and distance score as seen in Figure \ref{f:signal_ru}. Reference sequences usually are obtained in base representation (in the base equivalent `\textit{ACGT}' characters) and need to be converted to a synthetic signal representation before being used to map the reads. This can be done using the k-mer model which slides a window size of $k$ bases over the base reference while the bases in the window are mapped to a value using the k-mer model hash-map (see Figure \ref{f:signal_ru}). The obtained alignment position and score are then used to determine if a rejection should be made which is custom to application usage.
This signal-alignment workflow was first shown by Loose et al. \cite{loose2016real} in the \textit{RUscripts} work, which is also the first Read Until implementation introduced. \textit{RUscripts} is a Python implementation that uses the sDTW algorithm for aligning initial segments of the raw signals to the synthetic reference and was capable of matching 1 read every 0.3 seconds on a single CPU core \cite{loose2016real}. At the time of the proposal, \textit{RUscripts} was capable of keeping up with the 70 bases/s nanopore sequencing rate on a 22-core server \cite{loose2016real}. However, as sequencing speed improved over the years, the current 450 bases/s sequencing rate \cite{wang2021nanopore} has far surpassed \textit{RUscripts}'s capability of performing Read Until during sequencing. We observed that 98\% of processing time is spent processing the $O(MN)$ sDTW algorithm.

\textbf{Base-alignment Read Until.} As signal-aligning Read Until could not keep up with improved sequencing rates due to sDTW, researchers turned the focus of Read Until workflows towards base-domain techniques \cite{edwards2019real, payne2020nanopore}. These techniques align the genomic reads in the base domain as opposed to the signal domain which requires an extra step of basecalling the signal to base sequences in real-time prior to alignment (see Figure \ref{f:base_ru}). Thanks to well-optimized multi-state alignment implementations such as \textit{Minimap2} \cite{li2018minimap2} and the proprietary GPU-accelerated basecaller \textit{Guppy} from ONT, it is able to out-speed sequencing rate to save time. Yet, the extensive power usage and the need for high-performance GPUs and CPUs make base-alignment Read Until expensive and non-portable \cite{Stevanovski2022}. 

\textbf{Potential for signal-alignment Read Until.}
The above two alignment methods share high similarities in their algorithms and mainly differ in the sequence representation \cite{kruskal1983overview}. Since base-alignment Read Until are able to perform selective sequencing under current sequencing rates \cite{payne2020nanopore, Stevanovski2022} with the extra base-calling step, we revitalize the direct signal approach by optimizing and exploiting hardware acceleration for the sDTW alignment methodology targeting low-cost embedded heterogeneous platforms which also addresses the high cost of Read Until executions.

\subsection{Subsequence Dynamic Time Warping} \label{ss:bg-sdtw}
\begin{figure}[t] 
    \centering
    \begin{subfigure}{0.49\linewidth}
        \centering
        \vspace{1cm}
        \hspace{1cm}
        \includegraphics[width=0.9\linewidth,trim={0cm 0cm 0cm 0cm},clip]{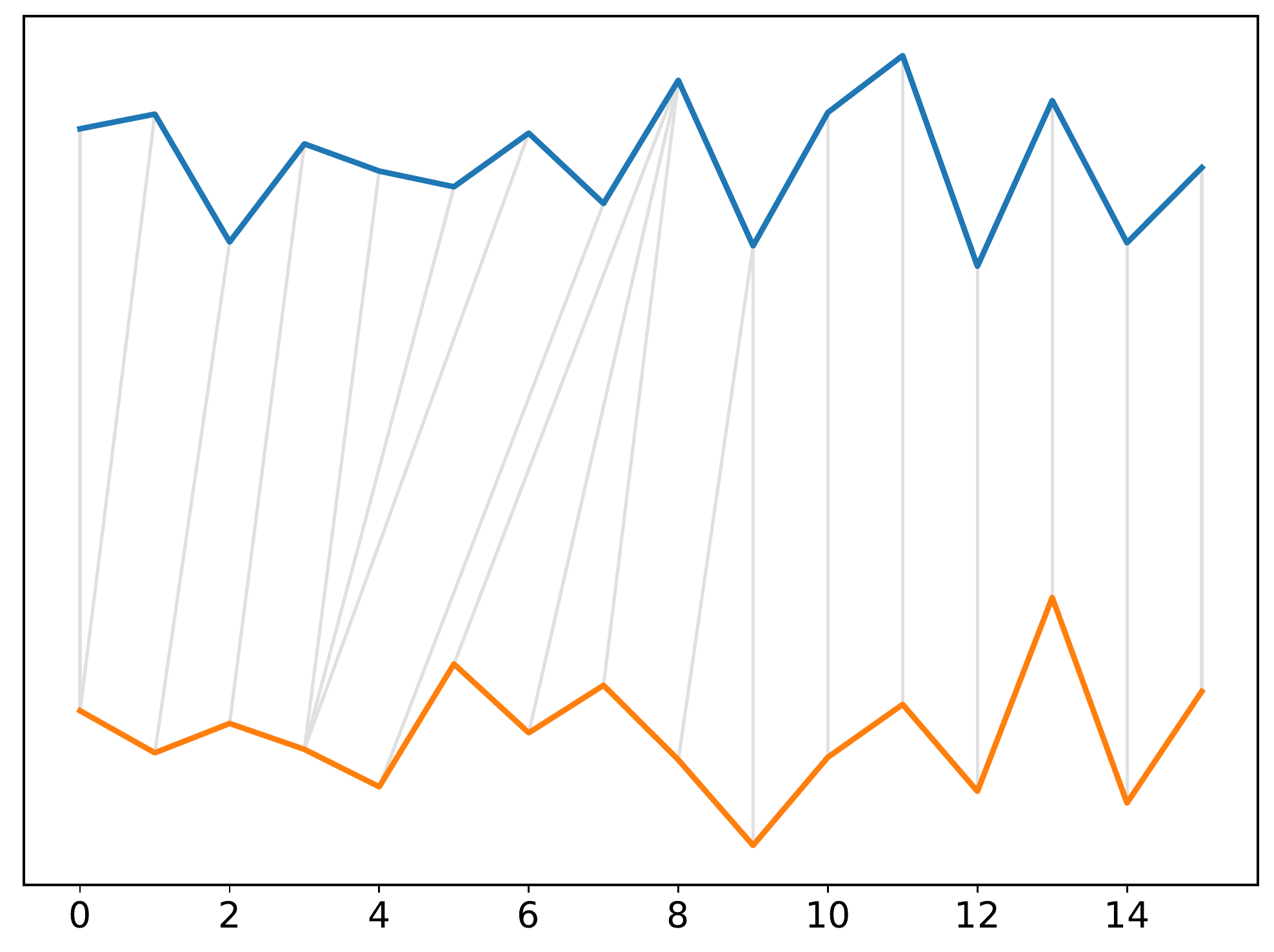}
        \caption{Signal warping} \label{f:dtw_warping}
    \end{subfigure}
    \begin{subfigure}{0.49\linewidth}
        \centering
        \includegraphics[width=\linewidth,trim={6cm 0.5cm 8cm 0cm},clip]{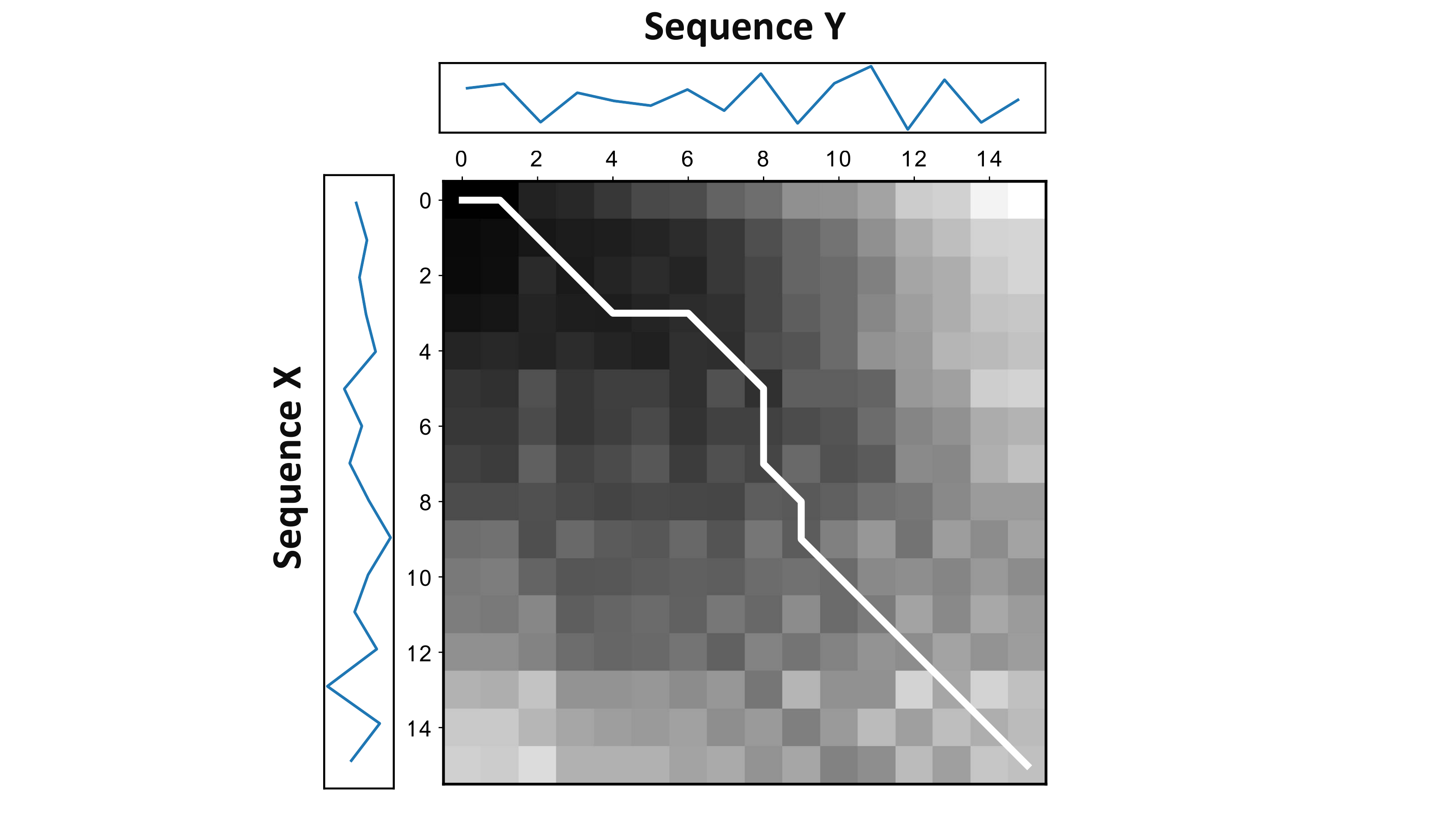}
        \caption{Classical DTW} \label{f:cdtw_matrix}
    \end{subfigure}\\
    \begin{subfigure}{0.99\linewidth}
        \centering
        \includegraphics[width=0.9\linewidth, trim={0.5cm 5.5cm 1cm 1cm},clip]{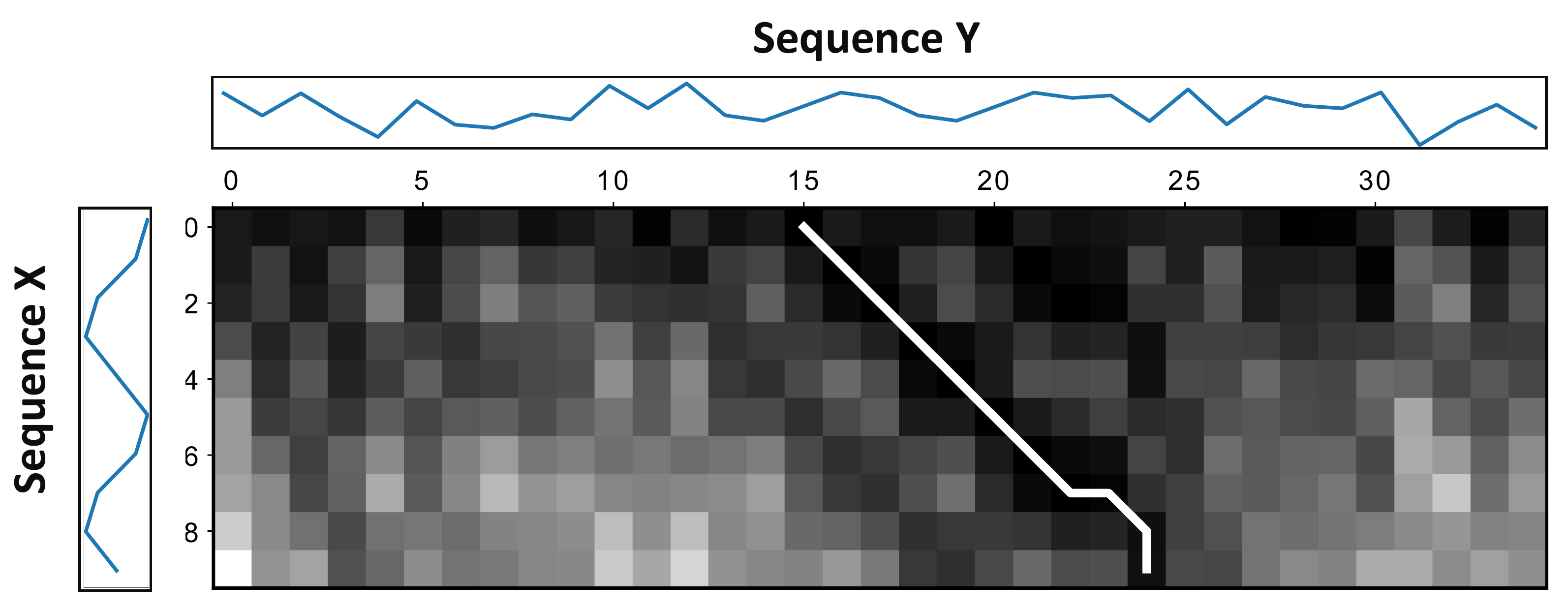}
        \setlength{\belowcaptionskip}{-1.5ex}
        \caption{Subsequence DTW} \label{f:sdtw_matrix}
    \end{subfigure}
    \setlength{\belowcaptionskip}{-1ex}
    \caption{Illustration of DTW}
    \label{f:dtw}
\end{figure}

The dynamic time warping algorithm family are dynamic programming algorithms that provide optimal alignment and distance metrics between two given time series \cite{muller2007dynamic} and have been widely used in pattern recognition applications in different fields \cite{juang1984hidden,tuzcu2005dynamic}.
This optimal alignment is achieved by warping the samples of the time series (see Figure \ref{f:dtw_warping}), which is done by keeping an $M\times N$ sized cost matrix. The classical DTW (cDTW) algorithm performs global alignment of the signals (see Figure \ref{f:cdtw_matrix}) \cite{muller2007dynamic} while the sDTW algorithm performs local alignment of the smaller sequence in the larger sequence (see Figure \ref{f:sdtw_matrix}) \cite{albanese2012mlpy}. 
Read Until attempts to find the local alignment of the query on the reference and thus utilizes sDTW which is elaborated below:



\textbf{sDTW Problem}: Given two sequences $X$ of size $M$ and $Y$ of size $N$ where $1\leq M\leq N \in \mathbb{N}$, the sDTW distance is the summation of the distance in the optimal warp path $w_{optimal}$.
The warp paths considered are all the paths that align the sequence $X$ with any subsequence of sequence $Y$.
The dynamic programming formulation of sDTW is based on the recurrence relation of equation:
\begin{equation} \label{eq:dtw_recurrence} \footnotesize
    \gamma(i, j) = \delta(i, j) + min
    \begin{cases}
    \gamma(i-1, j) \\
    \gamma(i-1, j-1) \\
    \gamma(i, j-1) \\
    \end{cases}
\end{equation}
where $\delta$ is the distance measure\footnote{Distance metrics in DTW are not limited to a single method. Popular distance metrics include Euclidean distance, squared Euclidean distance, and Manhattan distance.} between samples and $1\ge i\ge M$, $1\ge j\ge N$. The  boundary conditions for $\gamma(i,j)$ include $\gamma (i, 0)=\infty$ and $\gamma (0, j)=0$ and with a bottom-up memoisation, the $\gamma$ values are stored in a cost matrix $C$ of size $N\times M$ (i.e., $C[i, j]:= \gamma (i, j)$). $\gamma$ essentially chooses, at each step, the lowest cost move\footnote{In Equation \ref{eq:dtw_recurrence}, $\gamma(i-1,j)$ indicates an \textit{insertion} from sequence X into sequence Y whereas $\gamma(i-1,j-1)$ indicates a \textit{match} and $\gamma(i,j-1)$ indicates a \textit{deletion}.}. Once the cost matrix $C$ is populated, the cell with the minimum distance value in the last row would be the ending position of the local alignment. Backtracking from the end position by, again, choosing the step with the lowest cost among the same dependency will give the optimal warp path and starting position (see Figure \ref{f:sdtw_matrix}).

\setlength{\textfloatsep}{3ex}
\setlength{\floatsep}{3ex}

\setlength{\textfloatsep}{1ex}
\begin{algorithm}[t]
\footnotesize
\SetAlgoLined
\caption{Subsequence DTW} \label{alg:sdtw}
\SetKwInOut{Input}{Input}
\SetKwInOut{Output}{Output}
\Input{$X[1:M]$, $Y[1:N]$, $M$, $N$}
\Output{$position$, $score$}
$C$: cost matrix of size $N*M$\;
$score \gets \infty$\;
$position \gets -1$\;
\For{$j$ in range 1 to $N$} {
    $C[1, j] \gets abs(X[i]-Y[0])$\;
}
\For{$i$ in range 1 to $M$} {
    $C[i, 1] \gets abs(X[i]-Y[0]) + C[i-1, 1]$\;
}
\For{$i$ in range 1 to $M$} {
    \For{$j$ in range 1 to $N$} {
        $d\gets min(C[i-1, j], C[i, j-1], C[i-1, j-1])$\;
        $C[i, j]\gets abs(X[i]-Y[j]) + d$\;
    }
}
\For{$j$ in range 1 to $M$} {
    \If{$C[N, j] < min\_score$} {
        $position \gets j$\;
        $score \gets C[N, j]$\;
    }
}
\end{algorithm}

\textbf{Time and space complexity}: The sDTW approach is given in Algorithm \ref{alg:sdtw}. As shown, sDTW is $O(MN)$ in both time and space complexity due to the 2-dimensional search space. This has led to the heavy computational bottlenecks in applications such as \textit{RUscripts} discussed in Section \ref{ss:bg-sel_seq}. To date, there are not many sDTW optimization methods existing, and cDTW optimizations such as lower bounding \cite{keogh2006lb_keogh,lemire2009faster} and applying global constraints \cite{sakoe1978dynamic,itakura1975line} do not bring much benefit as the necessary search space is much larger than just the diagonal connecting start and end positions of the sequences.

\section{Related work} \label{s:related_work}
Existing hardware acceleration work targeting the subsequence search problem using the DTW algorithm family is rare. Previous FPGA accelerators such as \cite{sart2010accelerating,wang2013accelerating} implement a cDTW accelerator to compute the distance score between a query and a window buffer that stores a subset of the reference sequence. The reference sequence is continuously streamed into the window after each cDTW compute iteration completes which shifts older samples out. A distance score that is below a preset threshold indicates a match between the query and the current reference subsequence in the window buffer. \cite{sart2010accelerating} focused on exploiting coarse-grain parallelism by computing multiple cDTW in parallel. \cite{wang2013accelerating} introduced a PE-ring structure that computes multiple recurrence equations in parallel where the PEs compute cells that do not share data dependencies. This windowed cDTW approach is suitable for reference sequences of undetermined arbitrary length but is inefficient (requires $N\times$ $O(M^2)$ for software approaches) for the selective sequencing usage where the reference sequence is static with a known length. sDTW on the other hand is a data reusing version of the approach and our work exploits the fine-grain parallelism that computes the whole $O(M)$ dimension in parallel, leaving $O(M+N)$ computational time and $O(M)$ space.

For the hardware acceleration on signal-alignment Read Until, the only previous attempt was a simulated ASIC design that accelerates the sDTW algorithm \cite{michigan}. The proposed accelerator uses the unprocessed raw signal reads to directly map with the reference which requires 2000 PEs in total to perform the matching and has a reference limit of 100KB. The design has extensive resource requirements making it difficult to fit on lower-cost reconfigurable hardware, thus targeting ASIC. Furthermore, as seen in the history of Read Until \cite{loose2016real, edwards2019real, payne2020nanopore}, Read Until requires implementations to be fast adapting as nanopore sequencing technology fastly improves and the cost of re-manufacturing ASICs would be unsustainable.
In contrast, HARU is a fully complete design with an efficient software processing layer utilizing the sDTW accelerator. Our presented accelerator requires only 250 resource-efficient PEs due to pre-processing reducing the query size needed in the high-throughput computation of sDTW and is capable of executing selective sequencing with low-cost embedded MPSoC platforms with on-chip reconfigurable hardware.


\section{Hardware Accelerated Read Until} \label{s:haru}

\setlength{\textfloatsep}{2ex}
\begin{figure}[t] 
    \centering
    \includegraphics[width=0.95\linewidth,trim={1cm 0.5cm 10.5cm 0cm},clip]{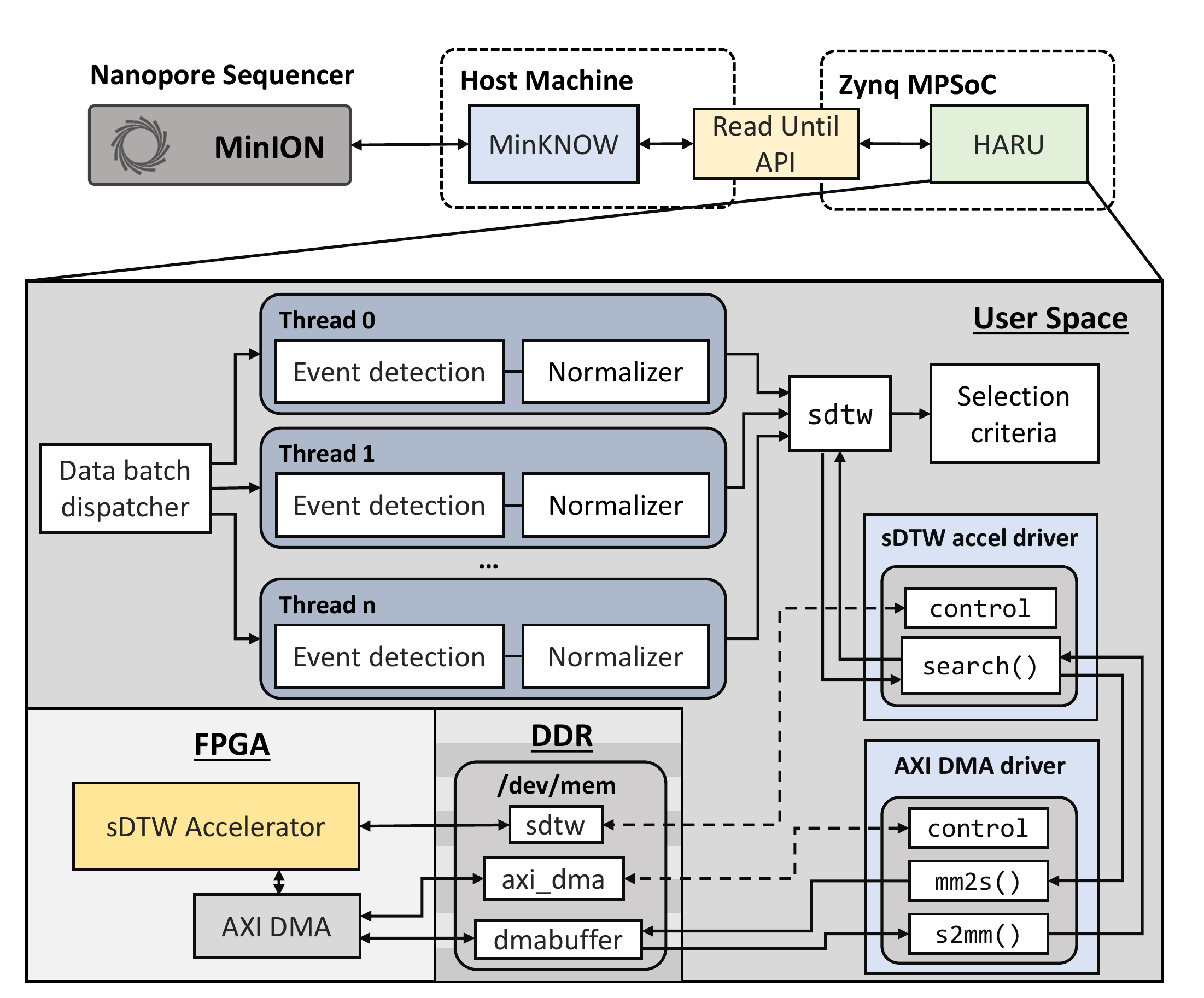}
    \setlength{\abovecaptionskip}{1ex}
    \caption{HARU architecture.} \label{f:haru_arch}
\end{figure}

HARU targets low-cost MPSoCs with on-chip FPGA to perform selective sequencing processing. Figure \ref{f:haru_arch} shows the architecture of HARU in an ONT nanopore sequencing workflow. HARU consists of three main components: the software processing layer, device drivers for the accelerator and associated hardware, and the hardware sDTW accelerator. The software processing layer, discussed in section \ref{ss:haru-software}, uses a multi-threaded batch processing architecture to perform raw read signal pre-processing and is customizable based on the selection criteria. The device drivers, discussed in section \ref{ss:haru-driver}, are designed to provide high-throughput data transferring of query and reference signals. Lastly, the resource-efficient sDTW accelerator, discussed in section \ref{ss:haru-sdtw_accel}, performs high throughput sDTW for the selective sequencing use case.

\subsection{Software Processing Layer} \label{ss:haru-software}
The software processing layer of HARU is the front-end of the HARU design running on the processing system on the MPSoC. Its main tasks include the pre-processing of the reference sequence and raw signal reads and the final selection decision. Since references are obtained in base representation as discussed in section \ref{ss:bg-sel_seq}, the initialization step of the software processing layer forms the synthetic reference signal for the forward and reverse representation of the base reference\footnote{This is needed since DNA molecules are double-stranded.} using the k-mer model for the flowcell type. Then, in preparation for the sDTW computation in hardware, the reference signal is normalized using z-score normalization. Since the data type used for the signal and cost matrix in the sDTW accelerator are 16-bit fixed-point types (discussed in section \ref{ss:haru-sdtw_accel}), the normalized values are scaled with a scaling factor to preserve signal resolution.

During the genome sequencing step, the software layer collects sequenced data from the nanopore sequencer in batches which is then dispatched into multiple threads for efficient computing of pre-processing (see Figure \ref{f:haru_arch}). Each thread performs event detection on the raw signal samples to reduce sample data size for the sDTW accelerator. This is done until enough events are collected. We find that for the R9.4 flowcell, 250 events are adequate for mapping and would require roughly 0.67 seconds of data collection\footnote{time to obtain around 50 events that belongs to the read adaptor and then actual 250 events of the query.}. After the collection, the events are normalized and scaled with the same scaling factor used in the reference signal preparation. When threads finish the pre-processing, the processed data are gathered and sent to the sDTW accelerator for processing using the drivers. After which, the mapping position and the similarity score are used to decide whether the read should be rejected.

\subsection{HARU Device Drivers} \label{ss:haru-driver}
To control and utilize the designed hardware accelerator in the software processing layer, we designed the software device drivers to have two main data paths (see Figure \ref{f:haru_arch}). The first data path is the control path of the accelerator which uses the AMBA AXI4-Lite protocol to configure the control registers and read status registers in software. The accelerator's physical address is memory-mapped to the virtual address space for user space applications to utilize.

The second data path is for the query and reference sequences. To prevent data transfer from becoming a bottleneck, we choose to use the AMBA AXI4-Stream protocol to stream query and reference data into the accelerator at a high-throughput rate. This is done by using AXI DMA module to point to a physical hardware address to stream data to and from. By calling the driver function for processing the query, the sDTW accelerator driver initiates the transferring from query and reference buffers to the transfer buffer on DDR memory dedicated for AXI-stream communication and to the FPGA.
Our benchmarks show that data can be sent to and from the accelerator at a throughput of 330 MB/s.

\subsection{Resource-Efficient sDTW Accelerator} \label{ss:haru-sdtw_accel}

\begin{algorithm}[t]
\footnotesize
\SetAlgoLined
\caption{Memory-efficient subsequence DTW}\label{alg:sdtw_mem_efficient}
\SetKwInOut{Input}{Input}
\SetKwInOut{Output}{Output}
\Input{$X[1:M]$, $Y[1:N]$, $M$, $N$}
\Output{$position$, $score$}
$C$: array of size $M + 1$ initialised to $\infty$\;
$score \gets \infty$\;
$position \gets -1$\;
\For{$j$ in range 1 to $N$} {
    $n \gets 0$\;
    $nw \gets C[1]$\;
    $w \gets C[2]$\;
    \For{$i$ in range 1 to $M$} {
        $C[i] := abs(x[i] - y[j]) + min(n, nw, w)$\;
        $n := C[i]$\;
        $nw := w$\;
        $w := C[i+2]$\;
    }
    \If{$C[M] < score$} {
        $position \gets j$\;
        $score \gets C[M]$\;
    }
}
\end{algorithm}

As discussed in section \ref{ss:bg-sdtw}, the standard sDTW algorithm has $O(MN)$ time and space complexity due to the computation of the cost matrix. The computation of a cell value in the cost matrix requires comparing three neighbor cell values, making the exploitation of available hardware parallelism harder. Also, preservation of the full cost matrix does not scale well if directly implemented on resource-constrained FPGA devices. We identified that the backtracking of the cost matrix to obtain the warp path is unnecessary for Read Until as the ending position is adequate to make the rejection decision. We provide the following optimizations over sDTW to obtain a resource-efficient high-throughput sDTW accelerator.

\begin{figure}[t]
    \centering
    \includegraphics[width=\linewidth,trim={1.2cm 3.5cm 7cm 3.6cm},clip]{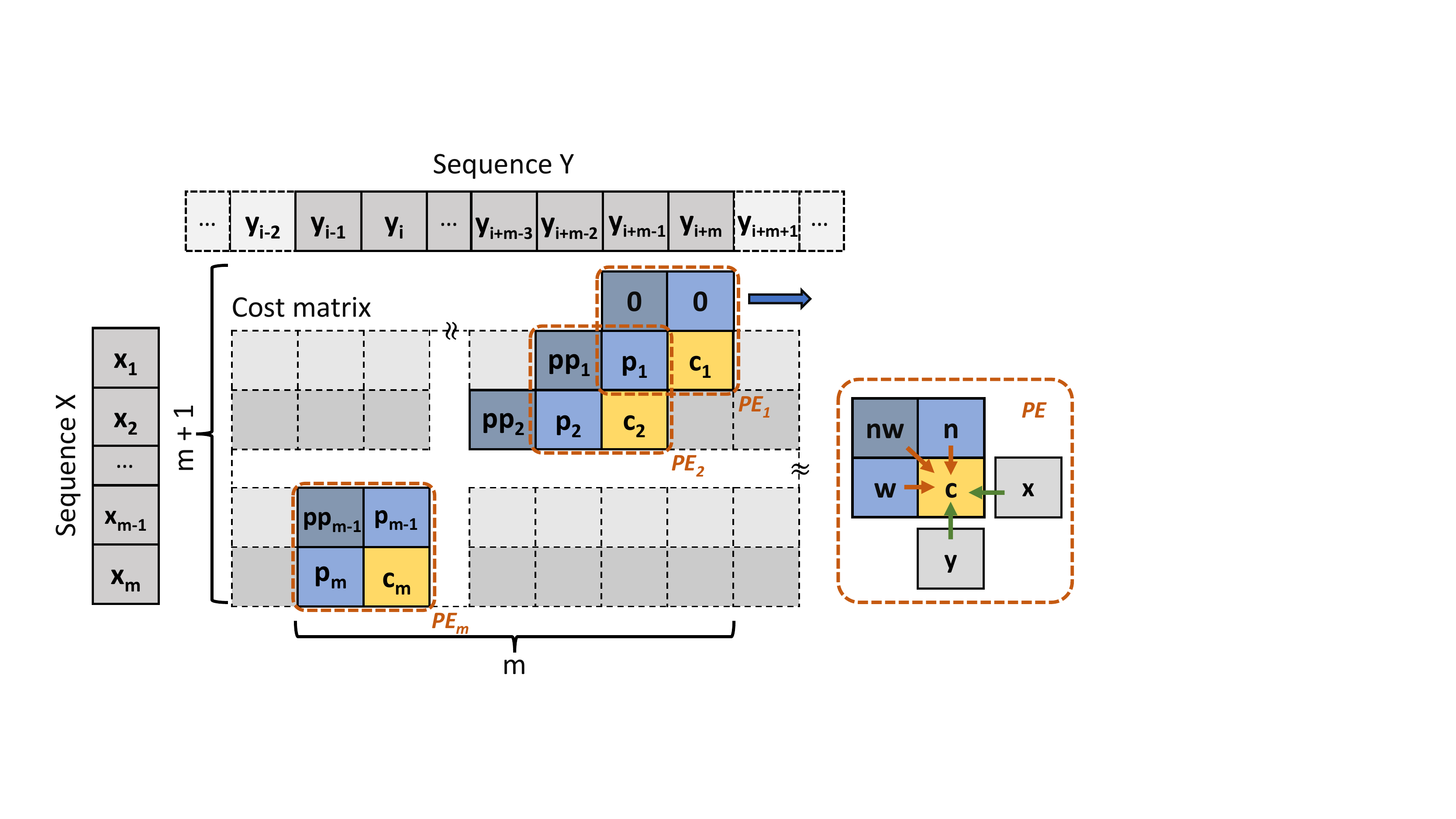}
    \setlength{\abovecaptionskip}{-1ex}
    \caption{Pipelined execution of Algorithm \ref{alg:sdtw_mem_efficient}} \label{f:sdtw_pipeline}
\end{figure}

\textbf{Cost matrix memory optimization.} The need to preserve the $M\times N$ sized matrix for backtracking was discussed in section \ref{ss:bg-sdtw}. However, for selective sequencing, the obtained end position of the alignment is adequate to determine the location of the current query, thus the backtracking step for obtaining the starting position is unnecessary. Consequently, the preservation of the whole cost matrix values is not necessary and a cost array of $M+1$ is sufficient. Algorithm \ref{alg:sdtw_mem_efficient} shows the sDTW algorithm after the cost matrix size is reduced. The outer loop (line 4 of Algorithm \ref{alg:sdtw_mem_efficient}) iterates through the whole reference sequence while the nested inner loop (line 8 of Algorithm \ref{alg:sdtw_mem_efficient}) iterates through the column at each reference sample. During each iteration of the inner loop, the computation of the recurrence equation is performed and the computed value is stored in the cost array that is of the same size as the query. Once the inner loop completes, the current minimum score and position values are updated if the last cell of the cost matrix is smaller than the current minimum score.

\textbf{Operation pipelining.} The sDTW cost matrix size reduction explained above optimizes the space complexity of the computation for selective sequencing. However, the execution of the algorithm is still sequential and has $O(MN)$ time complexity. Computing the whole column in parallel by unrolling the inner loop is not feasible due to the data dependency in the recurrence equation that needs waiting until the $n$ value is ready (see Algorithm \ref{alg:sdtw_mem_efficient}). We observe that once the first iteration of the inner loop for the column is completed, all data dependencies for the first inner loop iteration for the next column are ready. By pipelining the outer loop computation, an oblique column is formed, that is computed in parallel as shown in Figure \ref{f:sdtw_pipeline}. This oblique column traverses through the reference sequences, reducing the time complexity from $O(MN)$ to $O(N)$ since the $N$ query size is now computed in parallel.

\textbf{Fixed-point data representation.} After the optimisations explained above, the computational complexity in hardware is $O(M)$. However, the actual time needed is $(M+N-1)\times II$, where $II$ is the initiation interval (i.e. the number of cycles between loop iterations). In pipelined Algorithm \ref{alg:sdtw_mem_efficient}, $II$ is how fast the reference equation $C[i]:=abs(x[i]-y[j]) + min(n, nw, w)$ can be computed. Normally, 32-bit floating-point data types are used for the sDTW computation to preserve the precision after the sequences are normalized. This is expensive to implement in hardware in terms of resources and execution time. By using a fixed-point representation with fewer data bits and scaling the sequence values using a scaling factor, the recurrence equation can be computed in hardware rapidly and efficiently while keeping sufficient precision. We chose 16-bit fixed-point with a scaling factor of $2^5$ as it gives sufficient precision and keeps $II$ at 1 clock cycle (see section \ref{s:results} on accuracy). Note that, a 32-bit data type is used for the accumulation of distance score in cost arrays to avoid overflow. 



\begin{figure}[t]
    \centering
    \includegraphics[width=\linewidth,trim={5.85cm 1cm 5.3cm 1cm},clip]{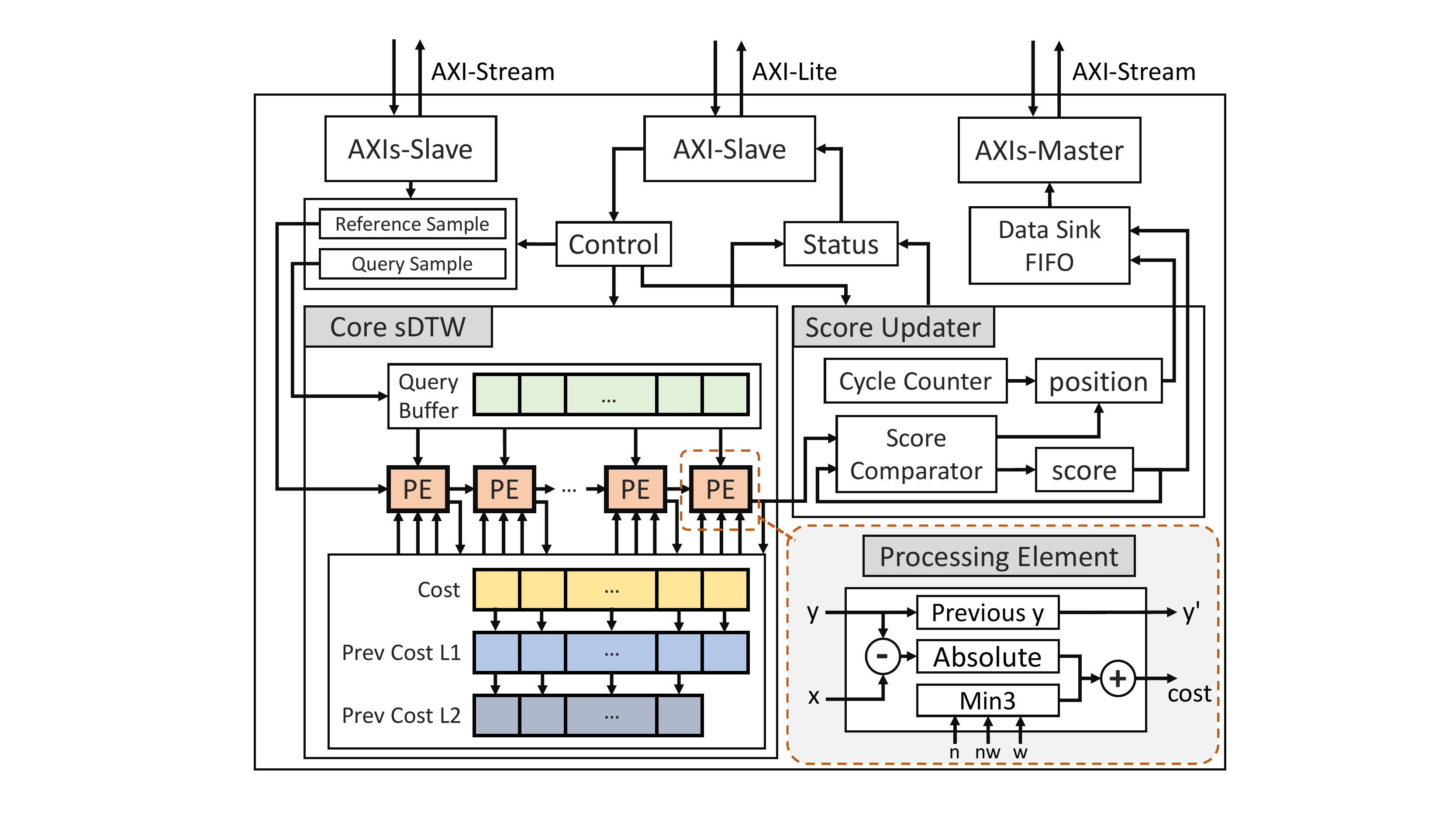}
    \setlength{\abovecaptionskip}{-1ex}
    \caption{sDTW hardware accelerator design for HARU} 
    \label{f:sdtw_accelerator}
\end{figure}

\textbf{HARU's sDTW Accelerator.}
The oblique parallel-computed column mentioned above is designed using a PE-chain structure where data dependent neighbor cells are shared amongst the PEs (Figure \ref{f:sdtw_pipeline}). As shown in Figure \ref{f:sdtw_accelerator}, the shared values are stored in two register arrays of size $M$ (L1 being the previous cost array and L2 being the second previous cost array). At each iteration, the costs in L1 array are shifted into the L2 array, while the current costs are passed onto the L1 array. Each PE computes the recurrence equation, which takes the Manhattan distance ($\delta=|x[i]-y[j]|$) and adds the minimum of the three neighbor cells (see equation \ref{eq:dtw_recurrence}). Samples of the reference sequence are first streamed into the first PE of the chain and are then passed along to successive PEs in each iteration. In section \ref{ss:haru-software}, we discussed that the software processing layer uses multi-threaded batch processing to perform event detection and normalization. The event detection decreases the size of the query to make the $M$ term smaller in the algorithm complexity. We choose to use a size of 250 events, giving the accelerator a PE chain of 250 PEs. In total, it takes $N+250-1$ clock cycles to complete the full search.



\setlength{\textfloatsep}{3ex}


\section{Experimental Setup} \label{s:experimental_setup}

The HARU system, proposed in section \ref{s:haru}, was implemented on a Xilinx's Kria AI Starter Kit that has a Zynq Ultrascale+ XCK26-SFVC784-2LV-C MPSoC. This board contains a processing system with a quad-core ARM Cortex A53 CPU and 4GB of DDR4 memory (specifications on column `MPSoC' in Table \ref{t:comp_platforms}). Implementation details of HARU will be discussed in section \ref{ss:exp-haru_implementation}. This HARU implementation is compared to two pure software implementations discussed in section \ref{ss:exp-pure_software}.
These two software versions are executed on a desktop computer comprising of a 10-core Intel Core-i9 processor and a high-performance computer (server) with a 36-core Intel Xeon processor (specifications are in Table \ref{t:comp_platforms}). We performed the experiments on two representative datasets detailed in section \ref{ss:exp-datasets}.



\setlength{\floatsep}{2ex}
\begin{table}[t]
\small
\centering
\caption{Computational platforms} \label{t:comp_platforms}
\vspace{-1em}
\begin{tabularx}{0.47\textwidth} { 
  >{\centering\arraybackslash}l 
  >{\centering\arraybackslash}X
  >{\centering\arraybackslash}X
  >{\centering\arraybackslash}X}
 \hline\hline
\textbf{System} & HPC & Desktop & MPSoC \\
\hline\hline
\textbf{CPU} & Intel Xeon Gold 6154 & Intel Core i9-10850K & Arm Cortex-A53 \\\hline
\textbf{CPU cores} & 36 & 10 & 4\\\hline
\textbf{Clock rate} & 3.00 GHz & 3.60 GHz & 1.5 GHz\\\hline
\textbf{RAM} & 377 GB & 32 GB & 4 GB\\\hline
\textbf{FPGA} & No & No & Yes \\\hline\hline\end{tabularx}
\end{table}

\begin{table}[t]
\small
\centering
\caption{Datasets} \label{t:exp_datasets}
\vspace{-1em}
\begin{tabularx}{0.47\textwidth} { 
  >{\centering\arraybackslash}l 
  >{\centering\arraybackslash}X
  >{\centering\arraybackslash}X}
\hline\hline
\textbf{Target} & SARS-CoV-2 & RFC1 \\\hline\hline
\textbf{Type} & Viral genome & Partial human genome \\\hline
\textbf{No. of bases} & 29,903 & 128,915 \\\hline
\textbf{Search space size} & 59,806 & 257,830 \\\hline
\textbf{No. of reads} & 1,382k & 500k \\\hline
\textbf{SLOW5 file size} & 5.5  GB & 39 GB \\\hline
\hline
\end{tabularx}
\end{table}

\setlength{\textfloatsep}{2ex}
\begin{figure*}[t]
    \begin{minipage}[c]{.66\textwidth}
        \centering
        \begin{subfigure}[!ht]{0.49\linewidth}
            \centering
            \includegraphics[width=\textwidth,trim={2cm 8.5cm 2cm 8.2cm},clip]{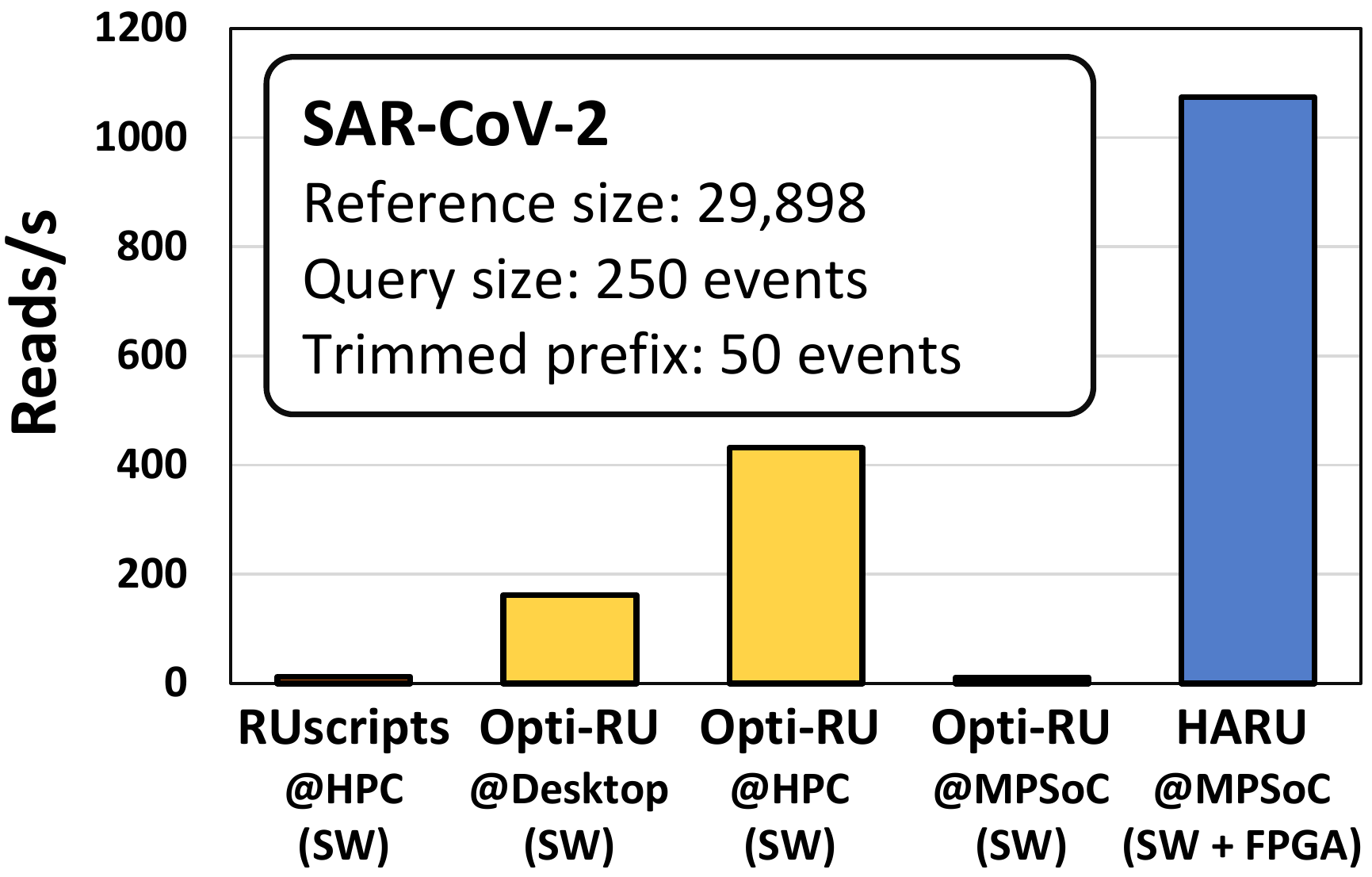}
            \caption{SARS-CoV-2 dataset} 
            \label{f:sp1_throughput}
        \end{subfigure}
        \begin{subfigure}[!ht]{0.49\linewidth}
            \centering
            \includegraphics[width=\textwidth,trim={2cm 8.5cm 2cm 8.2cm},clip]{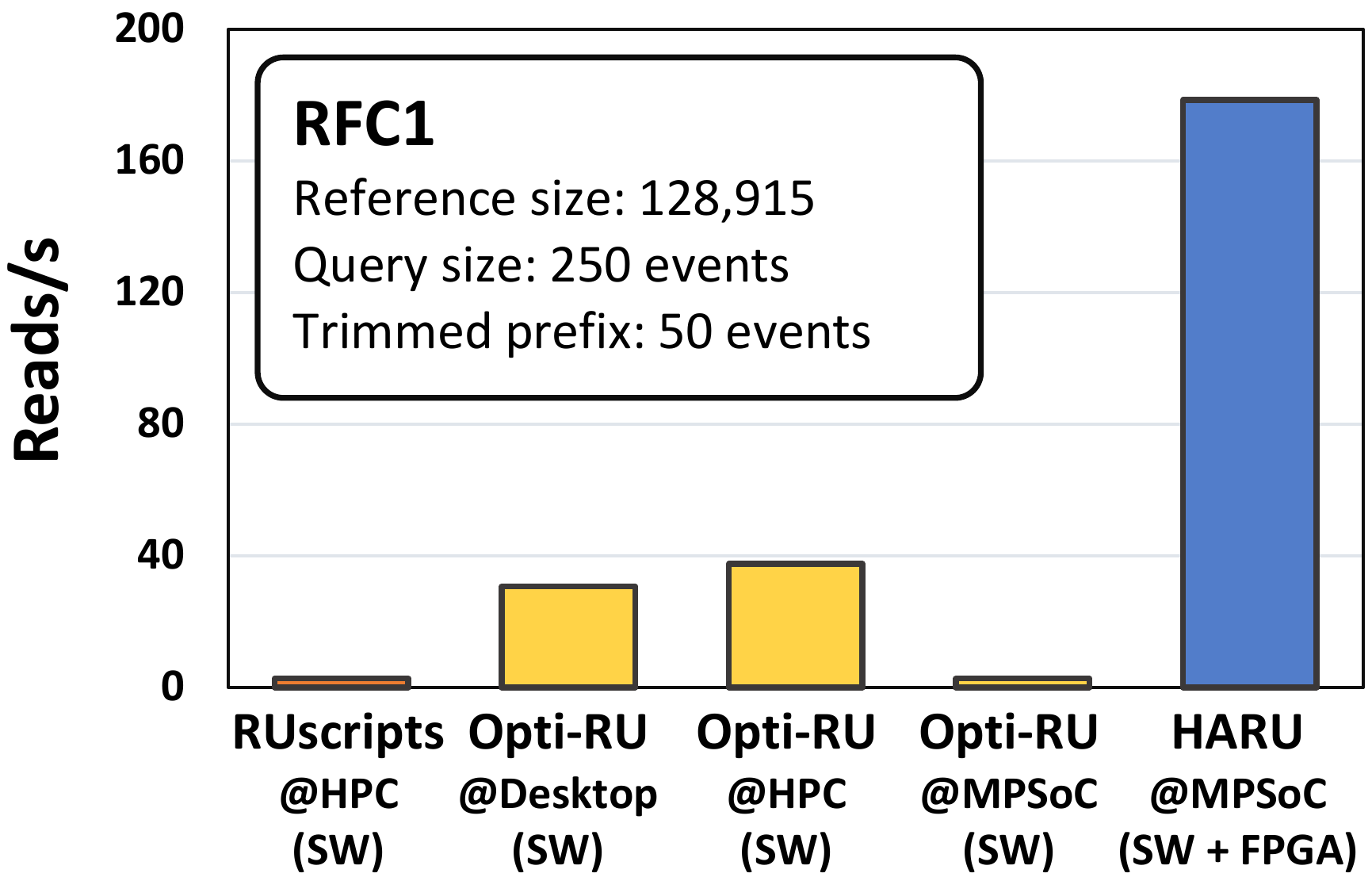}
            \caption{RFC1 dataset} 
            \label{f:rfc1_throughput}
        \end{subfigure}
        \label{f:haru_throughput}
        \setlength{\abovecaptionskip}{-0.005pt}
        \caption{Mapping throughput for the selective sequencing}
    \end{minipage}\hfill
    \begin{minipage}[c]{.33\textwidth}
        \centering
        \includegraphics[width=\linewidth,trim={2.3cm 3.4cm 3cm 4cm},clip]{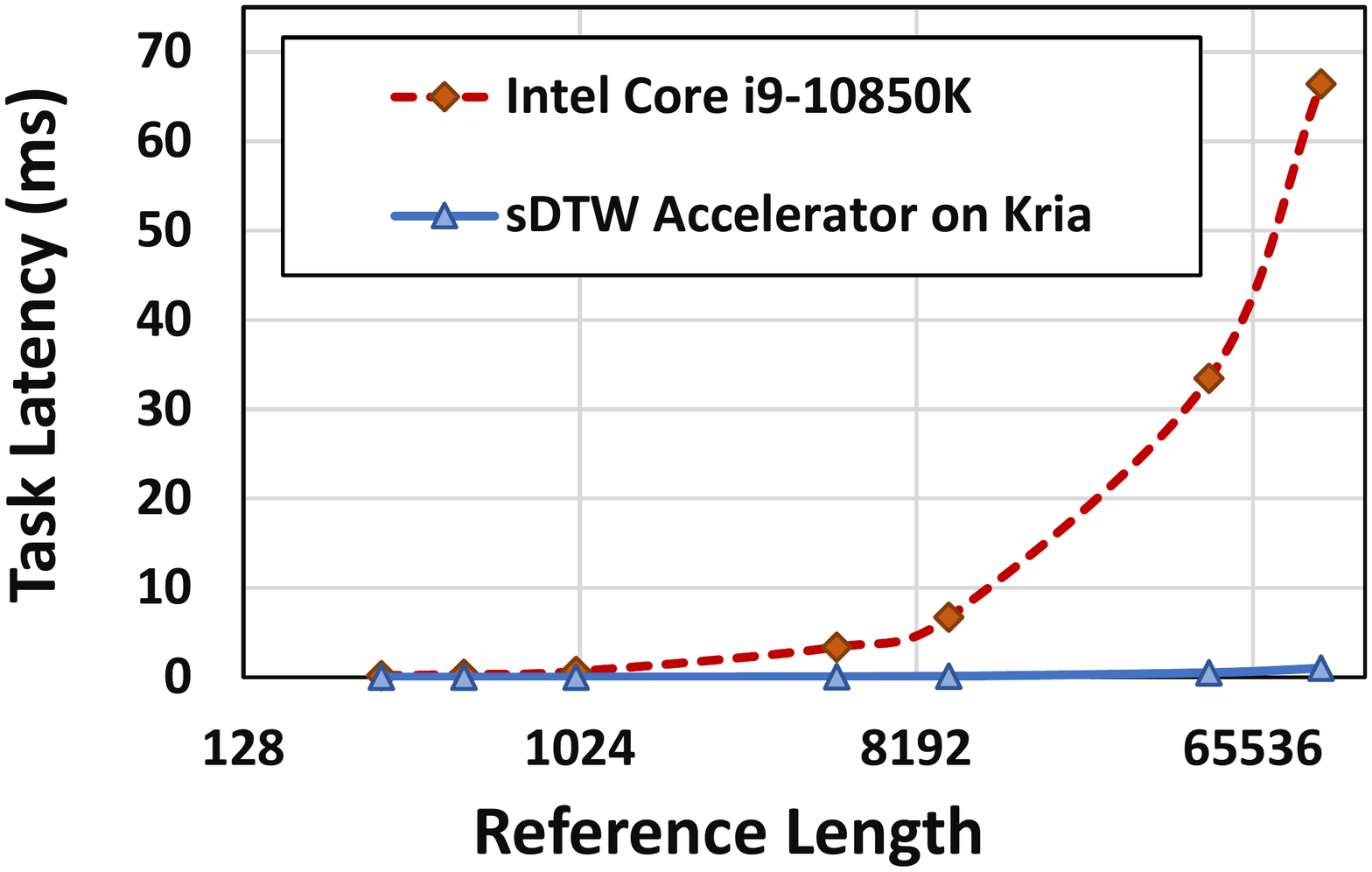}
        \setlength{\abovecaptionskip}{-0.01pt}
        \caption{sDTW task latency}
        \label{f:sdtw_throughput}
    \end{minipage}
\end{figure*}
\begin{figure*}[t]
    \begin{minipage}[c]{.33\textwidth}
        \centering
        \includegraphics[width=\linewidth,trim={2.5cm 3cm 2.1cm 2.9cm},clip]{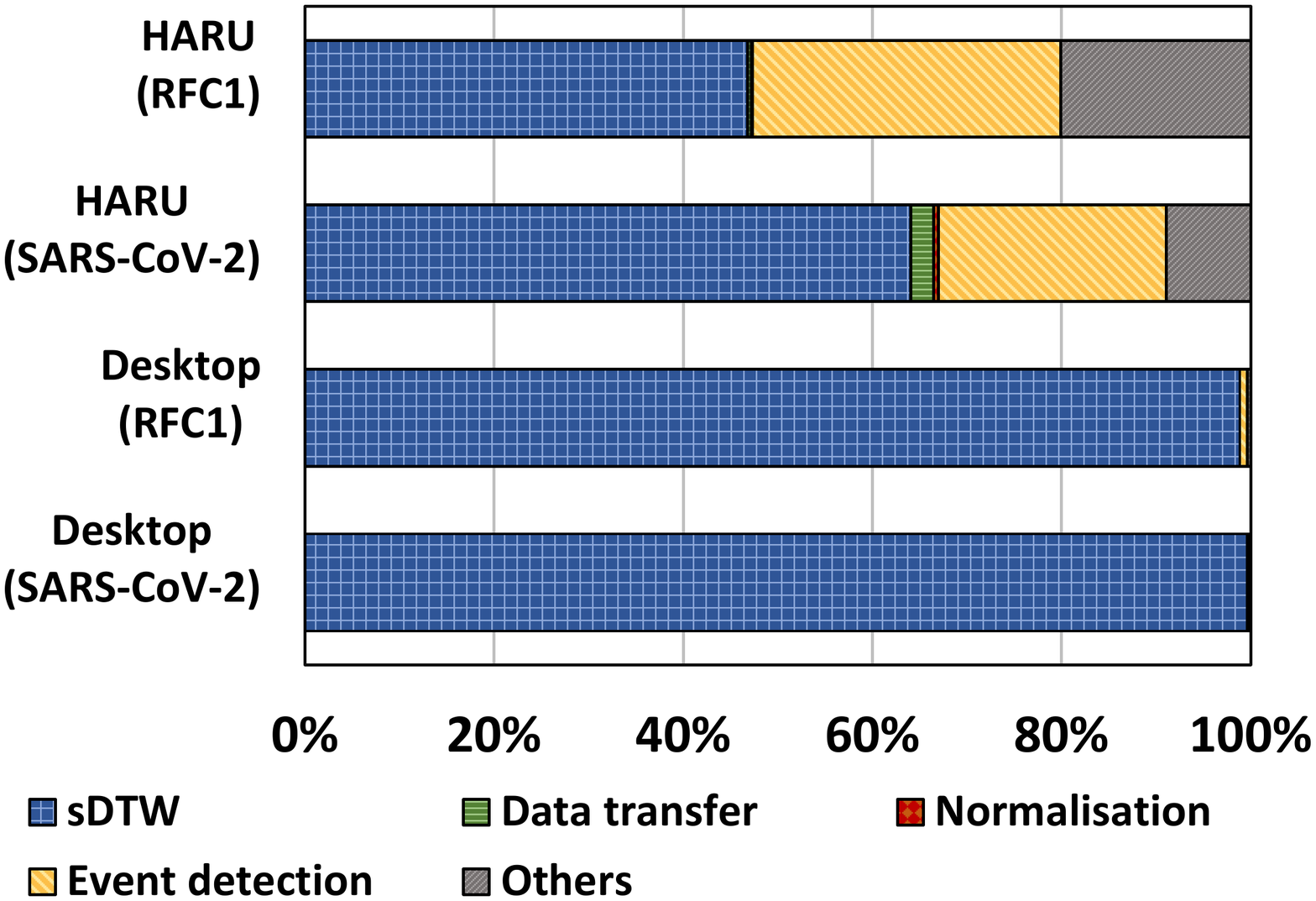}
        \setlength{\abovecaptionskip}{1.5ex}
        \caption{Process time breakdown} 
        \label{f:process_breakdown}
    \end{minipage}
    \begin{minipage}[c]{.33\textwidth}
        \centering
        \includegraphics[width=\linewidth,trim={2.9cm 3cm 3.5cm 3.4cm},clip]{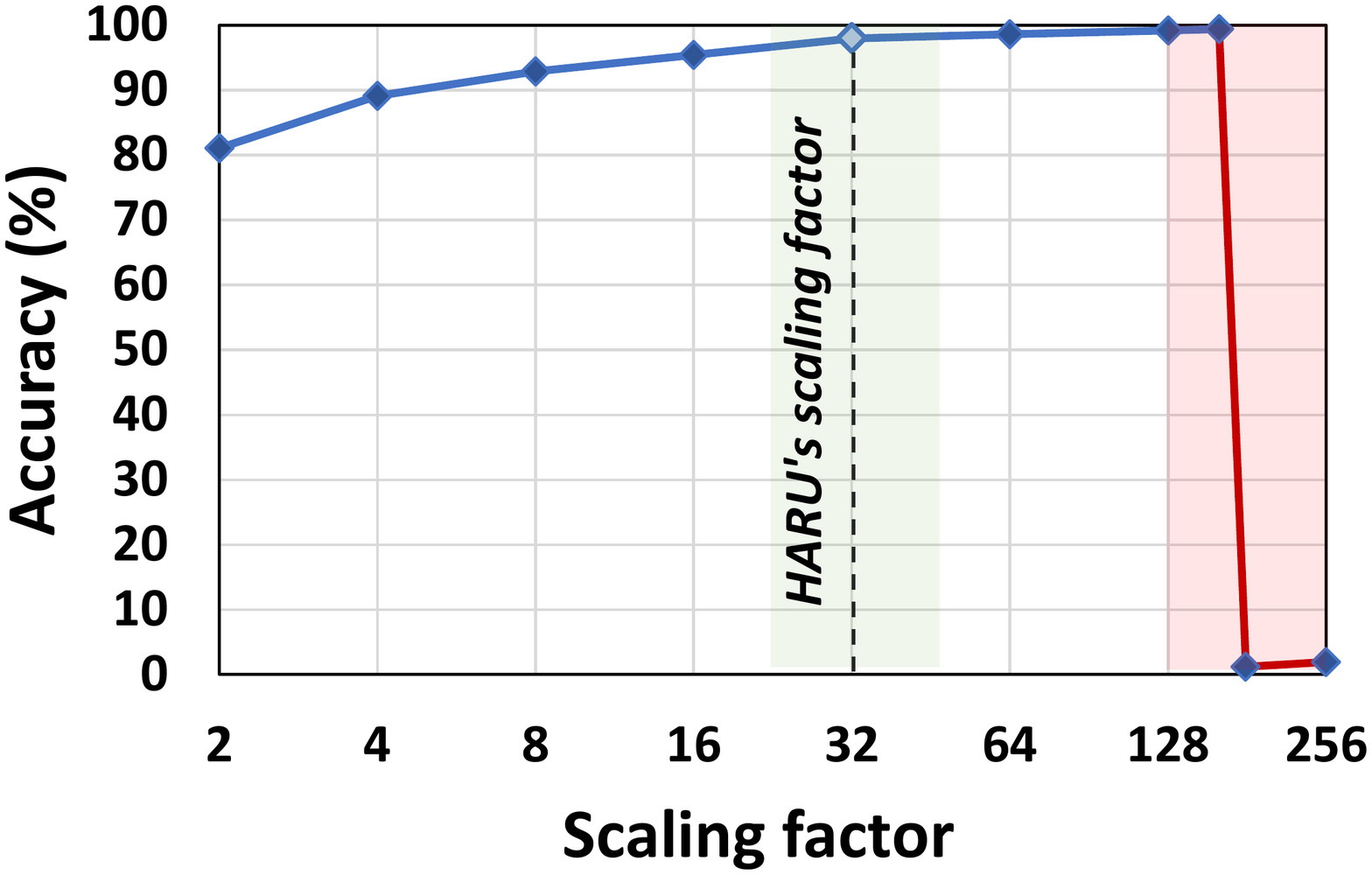}
        \caption{Accuracy against scaling factor}
        \label{f:scaling_accuracy}
    \end{minipage}
    \begin{minipage}[c]{.33\textwidth}
      \centering
        \vspace{1mm}
        \includegraphics[width=\linewidth,trim={2cm 2cm 2.3cm 1.9cm},clip]{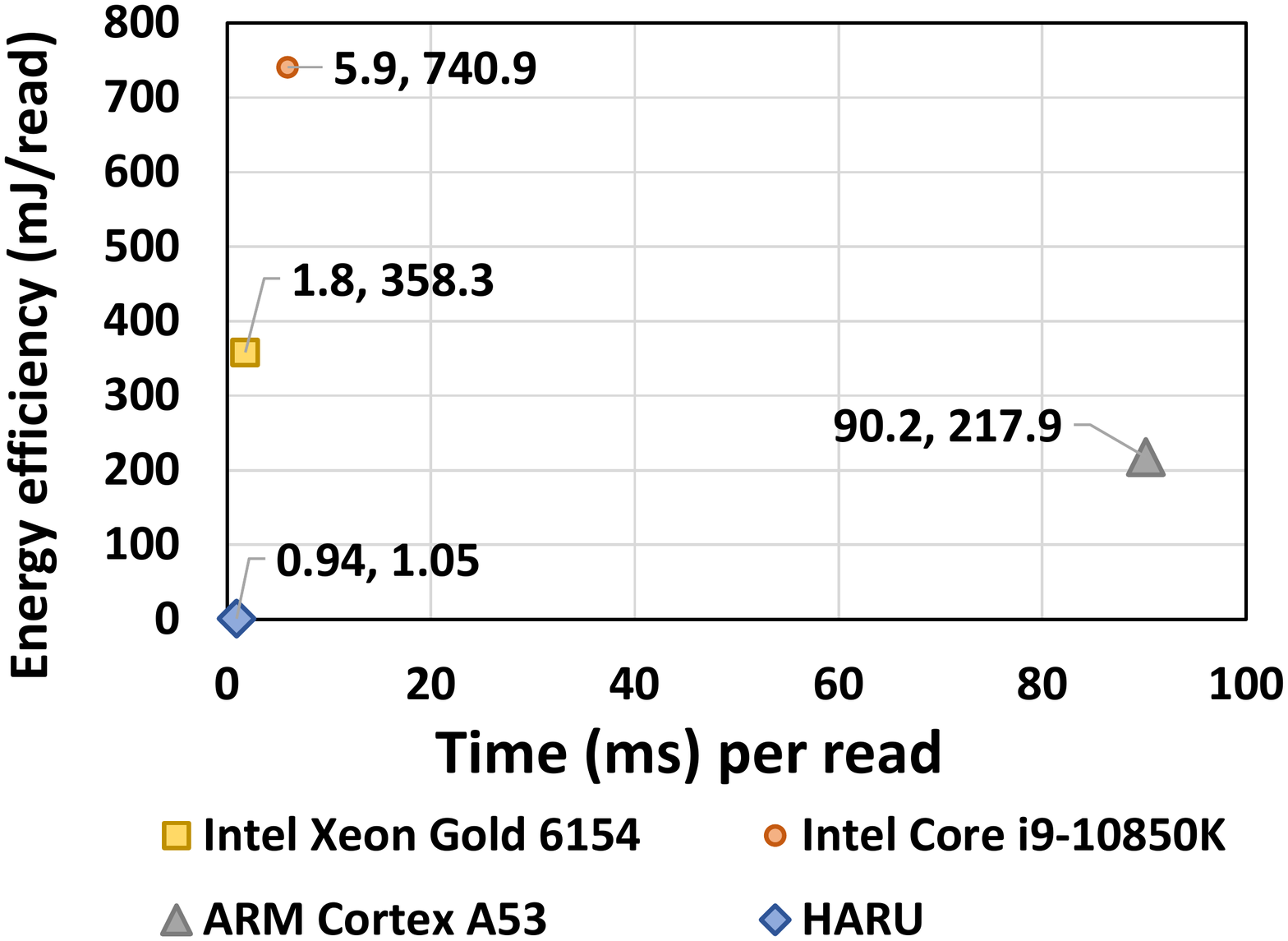}
        \setlength{\abovecaptionskip}{-2.2ex}
        \caption{Energy and performance} 
        \label{f:power_estimiates}
    \end{minipage}
\vspace{-1.8em}
\end{figure*}

\subsection{HARU implementation} \label{ss:exp-haru_implementation}
The operating system running on the processing system of the board is a customized embedded Linux image generated using Xilinx's Petalinux 2021.1 tool. To show the bare minimum throughput of the accelerator, our sDTW accelerator is synthesized with a single query processor in the accelerator running at 100 MHz. The number of query processors that can fit in the FPGA depends on the available resource on the device, see section \ref{ss:res-resource_utilization} for resource utilization of the accelerator with a single query processor.

\textbf{sDTW Hardware Accelerator}: The sDTW accelerator was implemented using Verilog HDL language. Synthesis was performed using Xilinx's Vivado 2021.1. The control bus interface for the accelerator utilizes the AMBA AXI-Lite protocol. For high-throughput data transfer for the query and reference data, we utilize the AMBA AXI-Stream protocol through the AXI DMA hardware in the FPGA.

\textbf{HARU Driver}: Device drivers were implemented for the hardware accelerator and AXI DMA in the C programming language. Both the accelerator and AXI DMA drivers memory map the physical address of corresponding devices into the virtual address space for utilization by the user space applications. The shared communication memory buffers between software and FPGA are preserved on the DDR memory which are allocated during the initialization stage.

\textbf{Software Processing Layer}: The software processing layer that prepares the raw signals and performs the selecting decision was implemented in the C programming language. For benchmarking experiments, the software loads raw signal data in the BLOW5 format \cite{slow5} from a USB 3 external hard drive connected to the Kria board. Raw signals for a batch of reads are first loaded to the RAM and are pre-processed using multiple threads implemented using POSIX threads. Pre-processing steps include event detection, prefix trimming, and normalization (explained in section \ref{ss:haru-software}). Then, sDTW is performed on each read on hardware by iteratively calling the HARU driver. Once the mapped positions and the DTW scores are available for the whole batch, the software computes the mapping quality (MAPQ) \cite{li2009sequence} and executes the selection criteria based on this score \cite{loose2016real}. 


\subsection{Pure software implementations} \label{ss:exp-pure_software}

\textbf{\textit{RUscripts}}: Original \textit{RUscripts} written by Loose et al. \cite{loose2016real} using Python 2.7 has reached end-of-life support and targets ONT's R7 Nanopore chemistry which is no longer in use. We modified \textit{RUscripts} to work on Python 3.6+ and extended it to support BLOW5 format and ONT's current Nanopore chemistry R9.4. This support for R9.4 chemistry is implemented by integrating the R9.4 pore-model and R9.4 event detection parameters from \cite{f5c,simpson2017detecting}.

\textbf{Optimised \textit{RUscripts} in C}: As the Python \textit{RUscripts} is not efficient enough for a fair comparison we implemented a multi-threaded C implementation that follows similar algorithmic steps. This implementation in C is very similar to the software explained above (section \ref{ss:haru-software}) except that sDTW on the CPU is called with multiple threads instead of using the FPGA accelerator. The sDTW computation used on the CPU is performed using the optimized sDTW implementation in the \textit{mlpy} library \cite{albanese2012mlpy}.  

\subsection{Datasets}\label{ss:exp-datasets}

HARU was tested against combinations of software running on the systems mentioned in Table \ref{t:comp_platforms} on two datasets. The first dataset is the SARS-CoV-2 genomic reads sequenced on a MinION flowcell and has a total of 1.382 million reads (Table \ref{t:exp_datasets}), publicly available at \cite{rodriguez2020covid}. SARS-CoV-2 genome (MN908947.3), which is 29903 bases long, is used as the reference for this experiment. The second dataset is a subset of a NA12878 human genome reference sample containing 500,000 reads sequenced on a PromethION flowcell (Table \ref{t:exp_datasets}), publicly available at \cite{slow5}. This dataset is mapped to a reference constructed by extracting the region chr4:39262456-39391375 (128 Kbases long) of the human genome (hg38). This region includes the RFC1 gene which contains an important pathogenic variant indicative of hereditary cerebellar ataxias disease and selective sequencing has been applied \cite{Stevanovski2022} for accurate diagnosis. 




\section{Results} \label{s:results}


\textbf{Overall system performance:} 
Figure \ref{f:sp1_throughput} compares the overall performance of HARU for mapping all the 1.382 million reads of the SARS-CoV-2 dataset with software-only implementations. The y-axis of Figure \ref{f:sp1_throughput} is the signal mapping throughput (mapping throughput is the execution time divided by the number of reads in the dataset). The First bar in Figure \ref{f:sp1_throughput} represents the original Python-based RUScripts run on the HPC with all 36 cores (throughput: 12.52 reads/s). The last bar represents our HARU system which has a throughput of 1073.83 reads/s). Thus, our HARU system is $\sim$85.8$\times$ faster than the original \textit{RUscripts}. The second bar shows the optimized C implementation of RUScripts (see section \ref{ss:exp-pure_software}) on the desktop system with a 10-core i9 processor and the throughput is 162.29 reads/s (HARU is 6.6$\times$ faster). Then, the third bar is for the optimized C implementation run with all 36 Xeon cores on the HPC and the throughput is 432.06 reads/second. HARU system being implemented on a low-cost embedded FPGA system is still $\sim$2.49$\times$ faster than the server. The fourth bar in Figure \ref{f:sp1_throughput} is for the optimized C implementation on the MPSoC run only on the 4-core ARM CPU which has a throughput of 11.09 read/second. Thus, HARU that utilizes the FPGA is 96.8$\times$ faster than running on the ARM processor alone. 


Similarly, Figure \ref{f:rfc1_throughput} compares the overall HARU performance for executing on all the 500,000 reads of the human dataset. HARU (last bar) is 64.5$\times$ faster than RUScripts on the 36-core HPC (first bar); 5.8$\times$ and 4.7$\times$ faster than optimized C implementation on the 10-core desktop (second bar) and 36-core HPC (third bar), respectively; and, 66.2$\times$ than the optimized C implementation on a 4-core ARM processor (fourth bar) alone.

Note that time measurement for the above throughput calculation for HARU includes all the overheads including reading signal data from the disk, raw signal pre-processing on software, and data transfer time to/from FPGA for HARU and our FPGA implementation is running at 100 MHz. The speedups observed for HARU over other systems in Figure \ref{f:sp1_throughput} (SARS-CoV-2 reference) are higher compared to those in Figure \ref{f:rfc1_throughput} (RFC reference) because the RFC reference is larger (128Kbases) than the SARS-CoV-2 reference (29Kbases) as explained below.


\textbf{Performance of the sDTW over reference length}:
Figure \ref{f:sdtw_throughput} shows how the performance of our sDTW core in HARU executed on the FPGA (including the overhead for data transfer to/from FPGA) and the pure software version of DTW executed on the CPU varies over the reference length. The X-axis is the reference length on a number of bases on the log scale. Y-axis is the time taken for a single sDTW query. For the CPU (red curve) where this y-axis represents the time for executing the sDTW function on a single CPU thread, whereas for the FPGA (blue curve) this is the time for processing on the FPGA plus the data transfer to and from the FPGA. Observe in Figure \ref{f:sdtw_throughput} how the gap between the two curves increases with the reference length, which causes the speed up of HARU over CPU to increase with increased reference size. This behavior is due to a band of cells being computed in parallel on hardware using a PE chain, as explained in section \ref{ss:haru-sdtw_accel}.

\textbf{Time breakdown for different processing steps:} Figure \ref{f:process_breakdown} compares the percentage of time spent on different processing steps for HARU vs the optimized software implementation in percentage. Due to the significant speedup of sDTW, the percentage of run time spent on sDTW is <64\% for the SARS-CoV-2 dataset and >46\% for the RFC1 data set (top two bars), whereas this was >98\% for software (bottom two bars). Note that `others' in Figure \ref{f:process_breakdown} is the time spent for loading data from the disk, reference preparation, and writing the output.

\textbf{Accuracy:}\label{ss:res-accuracy}
Figure \ref{f:scaling_accuracy} shows the accuracy of the accelerator using different scaling factors (discussed in section \ref{ss:haru-software}). Accuracy in Figure \ref{f:scaling_accuracy} is calculated as a percentage of the number of mapping positions that were similar to results produced from sDTW computed on software using  32-bit floating-point. Observe that a scaling factor of 2 yields a limited accuracy (80\%), while the increase of the scaling factor gradually converges the accuracy towards 100\%. However, when scaled above 128, the distance cost accumulation results in data overflow during sDTW which largely impacts the alignment accuracy. In HARU we have used a scaling factor of 32 to prevent overflow while having an accuracy close to 99\%.

\textbf{Energy comparison:} \label{ss:res-energy}
Figure \ref{f:power_estimiates} shows the estimated energy efficiency (y-axis) plotted against the execution time (x-axis) for HARU and optimized software-only implementations on different processors. HARU's overall performance and energy efficiency are considerably lower (close to the origin of the graph: time 0.94 ms/read and energy 1.05 mJ/read) than the optimized version running on ARM (90.2 ms/read, 217.9 mJ/read), Intel Core-i9 (5.9 ms/read, 740.9 mJ/read), and Intel Xeon Gold processor (1.8 ms/read, 358.3 mJ/read). The energy-delay product for the server is 644.94, where as, 0.987 for HARU, thus HARU is 650X better in terms of energy-delay product.
The energy consumed for HARU and the ARM processor was estimated by using the power estimates reported by Vivado in the synthesis report, whereas for Intel processors the Thermal Design Power (TDP) value reported in the processor specification was used.

\textbf{Resource utilization :} \label{ss:res-resource_utilization}
Resource utilization for our sDTW accelerator with a single query processor on Kria board as reported by the Vivado synthesis report is shown in Table \ref{t:resource_utilisation}. Note that for all the above experiments we used a single query processor to show the bare minimum performance on a low-end FPGA platform. As shown in Table \ref{t:resource_utilisation}, the maximum utilization (CLB LUT) is <20\% and thus the Kria board can support up to five parallel query processors if necessary.

\begin{table}[t]
\small
\centering
\caption{sDTW Accelerator resource utilization} \label{t:resource_utilisation}
\vspace{-1em}
\begin{tabularx}{0.47\textwidth} { 
  >{\centering\arraybackslash}l
  >{\centering\arraybackslash}X
  >{\centering\arraybackslash}X}
\hline\hline
\textbf{Resource} & Available & Used (utilisation) \\\hline\hline
\textbf{CLB LUT} & 117,120 & 21,570 (18.42\%) \\\hline
\textbf{CLB Registers} & 234,240 & 16,796 (7.17\%) \\\hline
\textbf{F7 Muxes} & 58,560 & 9 (0.02\%) \\\hline
\hline
\end{tabularx}
\end{table}

\section{Discussion} \label{ss:res-discussion}

In our proof-of-concept implementation of HARU, the reference sequence is first loaded onto the on-chip memory (block RAM) of the FPGA at the beginning of the execution. During alignment, the PE chain streams the reference samples from the block RAM to the first PE (Fig. \ref{f:sdtw_accelerator}). On-chip memory (block RAM) on the Xilinx Kria board used for evaluation is limited to 5.1 Mb, thus limiting the maximum reference sequence size to 295 kilobases. To eliminate this limitation, 
future work could directly stream the reference together with the query sequence, prior to each sDTW call (the query sequence is already streamed in the current HARU implementation). However, even with HARU (linear time complexity for sDTW), performing sDTW of a query on a giga-base-sized genome like the human genome is impractical (estimated to take ~3 seconds for a query). Nevertheless, when processing giga-based sized genomes, HARU is intended to be used in the final refinement step when potential mapping locations (a few reference sequence segments that are small in size) are first found using a heuristic method. Such a heuristic method that can map nanopore signals directly to giga-based sized genomes currently does not exist, yet, methods such as Sigmap \cite{zhang2021real} and UNCALLED \cite{kovaka2021targeted} are already setting the foundation for scalable direct signal mapping.

Future work can also focus on improving the throughput by implementing multiple parallel sDTW cores for coarse-grain parallelism. Our sDTW processor uses less than 20\% of the LUT resources of the FPGA as mentioned in Section \ref{ss:res-resource_utilization}. Thus, resources are sufficient to implement 5 parallel processors, which would increase the theoretical throughput by 5 times. A high-end FPGA board with a larger area could support even more processors, for instance, Xilinx's Versal VP2802 FPGA has sufficient resources to theoretically support ~140 parallel processors. However, such work also would require eliminating other bottlenecks that would arise. For instance, the 30\% of execution time currently spent on the signal pre-processing (Fig. \ref{f:process_breakdown}) would become a bottleneck then and will require acceleration.

Our implementation of HARU loads raw signal from BLOW5 file format because the slow5lib library itself is lightweight (with minimal dependencies), thus, easily allowing the cross-compilation to target the Kria platform. Running MinKNOW on the Kria platform is theoretically possible, but is far from practicality due to being closed source. Even if MinKNOW was open source, potential issues with hundreds of bulky dependencies will make cross-compilation impractical. Potential workarounds could include a server-client approach where MinKNOW runs on a laptop and communicates with the Kria board using ethernet. However, such workarounds are not ideal due to network communication overheads. Also, latency in the public-facing ReadUntil API provided by ONT (which is in Python programming language) would negate the massive benefit of having hardware acceleration. 

Our proof-of-concept HARU implementation is currently limited for DNA on R9.4 chemistry and future work could focus on extending for selective sequencing of RNA in future, or upcoming protein sequencing from ONT. Supporting the most recent R10.4 will be possible when a pore-model for R10.4 becomes available.

\section{Conclusion} \label{s:conclusion}
Existing sDTW-based software methods available for nanopore selective sequencing are highly computational intensive that a large workstation cannot keep up with a portable MinION sequencer. In this paper, we present HARU, a resource-efficient design that enables sDTW-based selective sequencing on a low-cost and portable heterogeneous system comprised of an ARM processor and an FPGA, which is around 85$\times$ faster than the original sDTW-based software implementation and around 2.5$\times$ faster than a highly optimized software version running on a server with a 36-core Xeon processor for a complete SARS-CoV-2 dataset. The energy-delay product for the server is around 650$\times$ higher than HARU executing on an embedded device. 


\section*{Source Code and Data Availability}

Source code for the HARU accelerator (including the Verilog HDL core accelerator and user-space device driver) is available at \url{https://github.com/beebdev/HARU}.  Source code that demonstrates the proof-of-concept integration of HARU accelerator for squiggle mapping is available at \url{https://github.com/beebdev/sigfish-haru}. The modified RUscripts to support Python 3.6+, BLOW5
format and ONT’s current Nanopore chemistry R9.4 is available at \url{https://github.com/beebdev/RUScripts-R9}.

Datasets used for the benchmarks are available to be directly downloaded from \url{https://doi.org/10.5281/zenodo.7314838}, which we curated from  publicly available datasets (\url{https://community.artic.network/t/links-to-raw-fast5-fastq-data-for-artic-protocol/17} associated with publication \cite{rodriguez2020covid} and 
 \url{https://www.ncbi.nlm.nih.gov/sra/SRX11368475} associated with publication \cite{slow5}).

\bibliographystyle{ACM-Reference-Format}
\bibliography{ref}

\clearpage
\end{document}